# Identification of a Damage Law

# by Using Full-Field Displacement Measurements

by

Damien Claire,[1,2] François Hild[1,*] and Stéphane Roux[2]

[1]LMT-Cachan

ENS de Cachan / CNRS-UMR 8535 / Université Paris 6

61 avenue du Président Wilson, F-94235 Cachan Cedex, France

[2]Laboratoire "Surface du Verre et Interfaces"

UMR CNRS/Saint Gobain 125

39 quai Lucien Lefranc, F-93303 Aubervilliers Cedex, France

[*]to whom correspondence should be addressed: Fax: +33 1 47 40 22 40

Email: hild@lmt.ens-cachan.fr

# Identification of a Damage Law by Using Full-Field Displacement Measurements

by

D. Claire, F. Hild and S. Roux

**Abstract:**

It is proposed to identify damage variables and their growth with loading in two dimensions by only using full-field displacement measurements. The equilibrium gap method is used to estimate the damage field during a biaxial experiment on a sample made of a composite material. From the analysis of a sequence of measurements, a proposed form of constitutive law is tested and identified. The emphasis is put on the methodology that can be applied to a vast class of materials.

# INTRODUCTION

The current development of reliable displacement field measurement techniques (Rastogi, 2000) allows for a better characterization of the behavior of materials and the response of structures to external loadings. Full-field measurements can be used in a variety of ways, namely:

- to check boundary conditions before performing the mechanical test itself (Calloch et al., 2001). In that case, it allows the experimentalist to control whether the boundary conditions correspond to the desired ones;

- to monitor an experiment (G'Sell et al., 1992, Fayolle, 2004) by using optical means as opposed to gauges or extensometers. In this area, one can envision using FE simulations as input "signals" to be compared with actual full-field measurements;

- to perform heterogeneous tests for which single measurements (e.g., by strain gauges, extensometers, clip gauges) are not sufficient to fully monitor an experiment, and particularly when the spatial heterogeneity is not known *a priori* [e.g., strain localization (Desrues et al., 1985, Bergonnier et al., 2005), damage localization (Berthaud et al., 1997) or crack initiation and/or propagation (Dawicke and Sutton, 1994, Forquin et al., 2004)];

- to study an experiment by using contactless techniques. This provides useful solutions to aggressive, hot, corrosive environments, or very soft solids for which gauges are not adapted [e.g., polymers (G'Sell et al., 1992, Chevalier et al., 2001), wood and paper (Choi et al., 1991), mineral wool (Hild et al., 2002, Roux et al., 2002)];

- to identify material properties. Identification techniques based upon the constitutive equation error (Ladevèze, 1975, Kohn and Lowe, 1988, Bui and Constantinescu, 2000) have been used in the determination of damage fields (Geymonat et al., 2002) or to study heterogeneous tests [e.g., Brazilian test (Calloch et al., 2002)]. Similarly, the so-called virtual fields method has been used to identify homogeneous properties of composites

(Grédiac, 1989, Grédiac et al., 2002, Grédiac, 2004) (i.e., in anisotropic elasticity). Another procedure is based upon the reciprocity gap (Bui, 1995) that can also be used to determine the local elastic field or to detect cracks in elastic media (Andrieux et al., 1997, 1999). In this paper, an alternative method solely based on displacement field data is used (Claire et al., 2002, 2004).

Continuum Damage Mechanics has reached a stage where the various forms of damage descriptions call for identification and validation in a systematic way (Allix and Hild, 2002). Various measurement techniques are used to evaluate damage variables (Lemaitre and Dufailly, 1987). The state coupling (Lemaitre and Marquis, 1992) between elasticity and damage (i.e., damage-induced loss of stiffness) will be used as a means of evaluating the damage state during an experiment. Instead of using single strain measurements as is usually performed (Lemaitre and Dufailly, 1977), full-field measurements are utilized. This requires resorting to non-conventional identification techniques. Contrary to conventional strategies where each test is designed to be homogeneous and thus where the sample behaves as a representative volume element, new approaches are now developed where a complex heterogeneous loading may allow the experimentalist in one single mechanical test to retrieve data about many different states (in terms of internal parameters) at once. Let us emphasize that this change is a "cultural" revolution whose consequences are expected to have a strong impact in a near future for experiments in solid mechanics.

To determine the mechanical properties of materials, inversion techniques are used as a means of identification or validation. When dealing with non-linear constitutive equations, one needs to postulate *a priori* the form of the constitutive equation to identify the unknown parameters (Calloch et al., 2002). It is proposed to analyze the distribution of damage and its change by using the so-called "equilibrium gap method" (Claire et al., 2004). For the sake of simplicity and computational effeciency, a simple isotropic damage model is introduced and used herein. The next section recalls the basic steps to perform an identification from displacement field measurements. The identification technique is then used to analyze a biaxial experiment on a composite material. First, from the displacement field, one determines the damage field for different load levels. Second, from the knowledge of the damage variable and its associated force, the corresponding growth is obtained without any *a priori* hypothesis on its dependence. A thermodynamic consistency as well as an acceptable identification error are then accounted for to refine the identification and a rescaling is proposed to compare damage fields determined independently from each analyzed load level. The particular challenge one faces here is to base the identification procedure on the *sole* use of kinematic measurements.

## DAMAGE MODEL

The analysis performed herein is based upon an isotropic damage description. For the sake of simplicity, only one damage variable $D$ is considered even though two can be needed (Burr et al., 1995). A Continuum thermodynamics setting is used (Germain et al., 1983). Under isothermal conditions, the material state is described by the infinitesimal strain tensor $\boldsymbol{\varepsilon}$ and the damage variable $D$ (with its usual bounds, namely, $D = 0$ for a virgin material and $D = 1$ for a fully damaged state) so that the state potential $\psi$ (i.e., Helmholtz free energy density) reads

$$\psi = \frac{1}{2}\boldsymbol{\varepsilon} : \mathbf{K}(D)\boldsymbol{\varepsilon}, \tag{1}$$

where ':' denotes the contraction with respect to two indices, and $\mathbf{K}$ the fourth order stiffness tensor that is written as (Lemaitre, 1992)

$$\mathbf{K}(D) = \mathbf{K}_0 \times (1 - D), \tag{2}$$

where $\mathbf{K}_0$ is the virgin stiffness tensor. In Equation (1), only a recoverable part of the state potential is considered. Consequently, it is assumed that no residual stresses are present, created or relaxed within the material during the whole load history. The associated forces are respectively defined by

$$\boldsymbol{\sigma} = \frac{\partial \psi}{\partial \boldsymbol{\varepsilon}} = \mathbf{K}_0 (1-D)\boldsymbol{\varepsilon} \quad \text{and} \quad Y = -\frac{\partial \psi}{\partial D} = \frac{1}{2}\boldsymbol{\varepsilon} : \mathbf{K}_0 \boldsymbol{\varepsilon}, \tag{3}$$

where $\boldsymbol{\sigma}$ is the Cauchy stress tensor, and $Y$ the energy release rate density (Chaboche, 1977). The fact that the following second derivatives of the state potential are different from zero

$$\frac{\partial^2 \psi}{\partial D \partial \boldsymbol{\varepsilon}} = \frac{\partial^2 \psi}{\partial \boldsymbol{\varepsilon} \partial D} \neq 0 \tag{4}$$

indicates a *state coupling* (Lemaitre and Marquis, 1992) between elasticity and damage. This coupling is used to measure indirectly damage variables by their influence on the stiffness variation [i.e., stiffness loss (Lemaitre and Dufailly, 1977, 1987)]. Clausius-Duhem inequality, in the present case, reduces to

$$Y\dot{D} \geq 0, \tag{5}$$

where a dotted variable corresponds to its first time-derivative. Since the energy release rate density $Y$ is a positive function, the damage growth is such that

$$\dot{D} \geq 0. \tag{6}$$

Within the framework of generalized standard materials (Halphen and Nguyen, 1975) and for a time-independent behavior, the damage growth can be written as (Marigo, 1981)

$$\dot{D} = \dot{d}\frac{\partial f}{\partial Y}, \tag{7}$$

where the damage multiplier $\dot{d}$ satisfies the Kuhn-Tucker conditions, and $f$ is the loading function. It follows that the damage growth for any load history can be recast as

$$D(t) = H\left[\max_{0 \leq \tau \leq t} Y(\tau)\right], \tag{8}$$

where $H$ is a monotonically increasing function to be identified. In the following, a heterogeneous test is used to determine the function $H$. Consequently, the damage distribution can no longer be assumed to be homogeneous. To identify the damage *field*, the equilibrium gap method is used since displacement measurements are available. In the sequel, the heterogeneity of the elastic field is reduced to a scalar and isotropic damage field $D(\mathbf{x})$. For this type of damage description, the Poisson's ratio is unaffected and the Lamé's coefficients can be written as $\lambda(\mathbf{x}) = \lambda_0 [1 - D(\mathbf{x})]$ and $\mu(\mathbf{x}) = \mu_0 [1 - D(\mathbf{x})]$, where the subscript 0 refers to reference quantities.

# THE EQUILIBRIUM GAP METHOD

An identification formulation is now presented in which the displacements $\mathbf{u}(\mathbf{x})$ are *known* and the elastic properties are unknown. This problem setting is unconventional in the sense that classical FE formulations assume known mechanical properties and try to determine the displacement field for different types of boundary conditions.

## Problem Setting

Let us consider a structure $\Omega$. When the considered medium is assumed to have damage discontinuities, a suitable setting is related to the equilibrium conditions corresponding to a continuity of the stress vector across a surface of normal $\mathbf{n}$

$$[\![\boldsymbol{\sigma}.\mathbf{n}]\!] = \mathbf{0}, \qquad (9)$$

where $[\![*]\!]$ denotes the jump of the quantity $*$. The jump conditions (9) are directly applied to a FE formulation. The potential energy theorem allows for a weak formulation of the equilibrium equations, which is linearly dependent on the displacements *and* elastic properties. Since most measurement techniques yield data on a regular mesh of points, the same hypothesis is made for the identification procedure. Consequently, quadratic square elements are considered for which each node corresponds to a measurement point. This hypothesis allows us to derive a specific formulation in which only middle nodes are considered. When the damage parameter $D_e$ is constant for a given element *e* occupying a domain $\Omega_e$, the elementary stiffness matrix $[\mathbf{K}_{me}]$ can be written as

$$[\mathbf{K}_{me}](D_e) = (1 - D_e) \times [\mathbf{K}_{me0}], \qquad (10)$$

where $[\mathbf{K}_{me0}]$ is the elementary stiffness matrix of an undamaged element [see Equation (2)]. Similarly, the strain energy $E_{me}$ can be written as

$$E_{me}(D_e) = \frac{(1 - D_e)}{2} \{\mathbf{u}_e\}^t [\mathbf{K}_{me0}] \{\mathbf{u}_e\}, \qquad (11)$$

where $\{\mathbf{u}_e\}$ is the nodal displacement column vector and $^t$ the matrix transposition. In the absence of external load on the considered nodes, the equilibrium conditions (9) can be rewritten for each middle node '12' of two neighboring elements 1 and 2

$$\frac{\partial E_{m12}}{\partial \mathbf{u}_{12}}(D_1, D_2) = \mathbf{0}, \qquad (12)$$

with $E_{m12}(D_1, D_2) = E_{m1}(D_1) + E_{m2}(D_2)$, where $D_1$, $D_2$ are the damage variables in elements 1 and 2, respectively. By writing this condition for each middle node, one ends up with a linear system in which the unknowns are the damage parameters assumed to be piece-wise constant and the known quantities are all the nodal displacements. In practice, Equation (12) is not strictly satisfied and a residual force $\mathbf{F}_r$ arises

$$\mathbf{F}_r(\hat{D}_1, \hat{D}_2) = \frac{\partial E_{m1}}{\partial \mathbf{u}_{12}}(\hat{D}_1) + \frac{\partial E_{m2}}{\partial \mathbf{u}_{12}}(\hat{D}_2), \qquad (13)$$

where $\hat{D}_1, \hat{D}_2$ are trial values of the unknown damage variables. The aim of the following section is to propose a practical setting for the identification of a damage field from the knowledge of displacement fields by minimizing the residuals $\mathbf{F}_r$. The method is therefore referred to as equilibrium gap method (Claire et al., 2004).

Practical Formulation

Since damage is assumed to be isotropic, a more appropriate setting can be used. Let us note $\{p_1\}$ the column vector of the nodal quantities of the first element and $\{p_2\}$ that of the second element. The $k^{\text{th}}$ equilibrium condition becomes

$$\overline{g}_k(\{\mathbf{p}_1\})(1-D_1) = \breve{g}_k(\{\mathbf{p}_2\})(1-D_2), \tag{14}$$

where $\overline{g}_k$ and $\breve{g}_k$ are generic functions depending on the nodal displacements (Claire et al., 2004). By considering all the equilibrium equations, the following global system is obtained

$$[\mathbf{G}]\{\mathbf{D}\} = \{\mathbf{g}\} \quad \text{with} \quad \{\mathbf{D}\} = \{D_1, D_2, ... D_N\} \tag{15}$$

where $[\mathbf{G}]$ and $\{\mathbf{g}\}$ are known and contain the nodal displacements. The advantage of this setting is that it can be written in a logarithmic form in which the displacements only appear in the right hand side of the following scalar expression

$$\ln(1-D_1) - \ln(1-D_2) = \ln|\breve{g}_k(\{\mathbf{p}_2\})| - \ln|\overline{g}_k(\{\mathbf{p}_1\})|. \tag{16}$$

Equation (16) automatically satisfies the requirement $D_e < 1$. However, this type of formulation can only be used for middle points. Corner nodes are concerned with four damage unknowns associated to the connecting elements and the same idea cannot be used. The system to solve is

$$[\mathbf{M}]\{\boldsymbol{\delta}\} = \{\mathbf{q}\} \tag{17}$$

where $\boldsymbol{\delta}$ is defined by

$$\{\boldsymbol{\delta}\} = \{\ln(1-D_1), \ln(1-D_2), ... \ln(1-D_N)\} \tag{18}$$

$[\mathbf{M}]$ is an assembled matrix corresponding to all the conditions (16) and $\{\mathbf{q}\}$ a vector that depends upon the nodal displacements. The system (17) is over-determined for the isotropic damage description used herein. For a structure $\Omega$, the following norm is minimized

$$\Gamma(\boldsymbol{\Delta}) = \|[\mathbf{M}]\{\boldsymbol{\Delta}\} - \{\mathbf{q}\}\|^2_{2(\Omega)} \tag{19}$$

with respect to $\boldsymbol{\Delta}$. A certain robustness can be expected thanks to the redundancy of the equations [e.g., for a square mesh made of $N$ elements, the number of equations $M$ is of the order of $4N$ (Claire et al., 2002)]. The minimization produces the following linear system

$$[\mathbf{M}]^t[\mathbf{M}]\{\boldsymbol{\Delta}\}=[\mathbf{M}]^t\{\mathbf{q}\} \qquad (20)$$

A variant to this formulation is to introduce a positive weight matrix $[\mathbf{W}]$ to modulate the different contributions in (19) according to the stress level. This is equivalent to modifying the norm $\|*\|_{2(\Omega)}$ and considering the norm $\|*\|_{W(\Omega)}$. The higher the stress vector, the higher the weight. In the linear system, we suggest to use $[\mathbf{W}]$ as a diagonal $M \times M$ matrix

$$[\mathbf{W}] = \begin{bmatrix} w_1 & 0 & \cdots & 0 \\ 0 & w_1 & \ldots & 0 \\ \vdots & \vdots & \ddots & \vdots \\ 0 & 0 & \ldots & w_M \end{bmatrix} \qquad (21)$$

with

$$w_k = \left| \bar{g}_k(\{\mathbf{p}_1\}) + \breve{g}_k(\{\mathbf{p}_2\}) \right|^\eta. \qquad (22)$$

The present choice consists in favoring equilibrium equations that are secure, because the load transfer between elements is large. This leads to the natural choice of decoupling different equilibrium equations, and hence to have a diagonal form for $[\mathbf{W}]$. The dependence of the magnitude of the weight $[\mathbf{W}]$ as compared to the magnitude of the load transfer remains the sole degree of freedom. Choosing a power-law relation is again an arbitrary choice. Finally, the power $\eta$ is adjusted so that the test cases with known damage distributions lead to the best results (Claire et al., 2002), a value $\eta = 1.5$ has been obtained. The system to solve becomes

$$[\mathbf{M}]^t[\mathbf{W}][\mathbf{M}]\{\boldsymbol{\Delta}\}=[\mathbf{M}]^t[\mathbf{W}]\{\mathbf{q}\} \qquad (23)$$

For Equation (23), the matrix $[\mathbf{M}]^t[\mathbf{W}][\mathbf{M}]$ has a zero eigen value and a corresponding eigen vector $\{\boldsymbol{\Delta}\}^t = \{1,1,\ldots 1\}$ (i.e., this corresponds to a global rescaling of the local elastic constants, or the $(1 - D)$ field, by a fixed multiplicative factor which does not affect the solution). Consequently, one can arbitrarily set one damage component of $\{\boldsymbol{\Delta}\}$. For simplicity, let us choose the $i_0^{\text{th}}$ component and consider the following initial condition

$$\{\boldsymbol{\Delta}\}_0^t = \{0, 0, \cdots, 0, \ln(1-D)_{i_0}, 0, \cdots, 0\} \tag{24}$$

and

$$\{\boldsymbol{\Phi}\} = \{\boldsymbol{\Delta}\} - \{\boldsymbol{\Delta}\}_0 \tag{25}$$

one needs to solve over the $(N-1)$ degrees of freedom of $\{\boldsymbol{\Phi}\}$, $i = 1,...,N$ and $i \neq i_0$. This corresponds to omitting the $i_0^{\text{th}}$ line and column of the $[\mathbf{M}]^t[\mathbf{W}][\mathbf{M}]$ matrix. The zero-eigenvalue is unique and thus the resulting matrix is now positive definite

$$[\mathbf{M}]^t[\mathbf{W}][\mathbf{M}]\{\boldsymbol{\Phi}\} = [\mathbf{M}]^t[\mathbf{W}]\{\mathbf{q}\} - [\mathbf{M}]^t[\mathbf{W}][\mathbf{M}]\{\boldsymbol{\Delta}\}_0. \tag{26}$$

This linear system can be solved by using different numerical methods. A conjugate gradient technique (Press et al., 1992) making use of the sparseness of the matrix $[\mathbf{M}]^t[\mathbf{W}][\mathbf{M}]$ is utilized. When artificial measurements are used, a comparison could be performed with an a priori prescribed damage field. An overall quality of the order of a few percents is achieved in all the configurations tested (Claire et al., 2002, 2004). When some additional noise was considered, the error did not change significantly.

Error Estimator

From Equation (15), an error indicator can be defined when the exact solution is unknown

$$\kappa = \|[\mathbf{G}]\{\mathbf{D}\} - \{\mathbf{g}\}\|_{2(\Omega)} \tag{27}$$

The quantity $\kappa$ characterizes the average equilibrium residuals. From this point of view, it is close to the indicator based on equilibrium residuals used to assess the quality of a FE computation (Babushka and Rheinboldt, 1978b, 1978a, Zienkiewicz and Taylor, 1988).

However, a simple dimensional analysis shows that $\kappa$ depends on the stress scale, which is exogenous to the present problem (i.e., based solely on kinematic measurements). Therefore, even though the solution is defined up to a constant scale factor in $(1 - D)$, $\kappa$ *does* depend on that factor. Consequently, the absolute scale for $\kappa$ is meaningless. Only relative values can be utilized (Claire et al., 2004). This is a central difficulty one faces in the sequel. It results from the choice of making use of kinematic data alone.

Because of the stress scale sensitivity, it is of interest to introduce another quantification of the suitability of a numerical solution to the identification problem. Associated to the $n^{\text{th}}$ middle node where the residual force is $\mathbf{F}_r(D_1,D_2)$, the work $W_r(D_1,D_2)$ is defined as

$$W_r(D_1, D_2) = \left| \mathbf{F}_r(D_1, D_2) \cdot \tilde{\mathbf{u}}_{12} \right|, \tag{28}$$

where the chosen displacement $\tilde{\mathbf{u}}_{12}$ is the measured displacement vector from which the rigid body motion of elements 1 and 2 has been removed. The magnitude of $W_r(D_1,D_2)$ can be compared to the elastic energy $E_{me12}(D_1,D_2)$ in the two considered elements 1 and 2 so that the following local indicator $\theta$ no longer depends on the unknown stress scale

$$\theta(n) = \frac{W_r(D_1, D_2)}{E_{me12}(D_1, D_2)}, \tag{29}$$

An error per element can be defined as

$$\theta_e = \sum_{n=1}^{n_m} \theta(n), \tag{30}$$

where $n_m$ is the number of middle points for the considered element $e$ (i.e., generally 4 for an inner element, 3 for edge elements and 2 for corner elements, since the boundary conditions in terms of load are not considered). A global indicator $\Theta$ can also be defined

$$\Theta = \frac{W_r}{E_m}, \tag{31}$$

where $W_r$ is the total work done by the residuals and $E_m$ the corresponding total elastic energy. The last indicators are *independent* of the stress scale factor. They will be used in the following to analyze a biaxial experiment on a composite material.

# DETERMINATION OF A DAMAGE LAW BY ANALYZING A HETEROGENEOUS EXPERIMENT

A vinylester matrix reinforced by E glass fibers is studied (Figure 1-a). A quasi-uniform distribution of orientations leads to an isotropic elastic behavior prior to matrix cracking and fiber breakage, which are the main damage mechanisms (Collin et al., 1998). A cross-shaped specimen is loaded in a multiaxial testing machine (Figure 1-b). The experiment is performed in such a way that the forces applied along two perpendicular directions are identical. Their norm is denoted by $F$. The displacement field of Figure 1-c is measured by digital image correlation. Each "measurement point" corresponds to the center of an interrogation window of size $64 \times 64$ pixels, equivalent to a surface of about 8 mm$^2$. At this scale, the material is not homogeneous (see Figure 1-a). The shift between two neighboring measurement points is 32 pixels. A sub-pixel algorithm is used. It enables for a displacement resolution of one hundredth of one pixel for 8-bit pictures studied herein. To achieve a better robustness, a hierarchical multi-scale version was used (Hild et al., 2002).

Five different load levels are analyzed, namely, $F = 5$ kN, 7 kN, 9 kN, 10 kN, and 11 kN. For each load level, $19 \times 23$ displacement measurements are obtained, from which $9 \times 11$ values of damage are evaluated. Figure 2 shows the $(1 - D)$-field computed from the measured displacement field. From the analysis of Figure 1-c, a crack clearly appears on the top left corner for the last load level before failure. This crack can be observed by the three dark elements. For the three first load levels, one can note at least three different corners where the damage value becomes significant. At this stage, crack inception is likely to have occurred in these three corners. One of them subsequently became preeminent as can be seen on the last load level. This type of analysis cannot be performed by only looking at the displacement field measurements. It shows that the present approach is able to give additional ways of analyzing experimental measurements. In Figure 3, the change of the error indicator $\Theta$ with the applied load $F$ is shown. Up to approximately 9 kN, the quality of the identification is identical. It starts to degrade for 10 kN and strongly changes for 11 kN, thereby indicating a change that can be attributed to macrocrack initiation.

In addition to $1-D$, the thermodynamic force $Y$ under plane stress assumption is computed from the in-plane strain field in a non-dimensional way

$$\frac{2Y}{E_0} = \frac{\varepsilon_{11}^2 + 2\nu_0\varepsilon_{11}\varepsilon_{22} + \varepsilon_{22}^2}{1-\nu_0^2} + \frac{2\varepsilon_{12}^2}{1+\nu_0}, \tag{32}$$

where the directions 1 and 2 are associated to an in-plane frame, $E_0$ the Young's modulus of the virgin material and $\nu_0$ the corresponding Poisson's ratio (here taken equal to 0.28). The damage growth is written in terms of $1-D$ versus $2Y/E_0$, which will be referred to as dimensionless strain energy release rate density. Figure 4 shows the changes for the five load levels. It should be remembered that upon performing the identification, no damage growth is assumed. Consequently, a scatter is to be expected. To analyze the whole sequence, the first convention that was chosen is to set the maximum value of $1-D$ to unity (i.e., $D=0$) for *each* load level. To allow for a multiplicative correction, the following damage kinematics is assumed

$$1-D = \min\left[1, A(F)\left(\frac{2Y}{E_0}\right)^\alpha\right], \tag{32}$$

where the parameter $A$ can be load-dependent (since $1-D$ is determined up to a multiplicative constant), and $\alpha$ a constant power. For all the load levels, the power $\alpha$ only slightly varies around the $-0.37$ value. Conversely, as anticipated, the prefactor $A$ is load-dependent as shown in Figure 4. The quality of the identification is assessed by computing the RMS error associated to the (1 − D)-field; in the present case, it is equal to 0.08.

The second step consists in performing a coupled identification by prescribing the *same* value for the power $\alpha$, and keeping the parameter $A$ different for each load level. A value for $\alpha$ equal to −0.37 is found. By assuming that the smallest damage level (for the lowest load level) is equal to 0, a correction is performed to 1 − D so that the prefactor of the damage growth law is identical (Figure 5) for all the Y range. From the present analysis, a *single* growth law is obtained that is valid over more than three decades of the dimensionless energy release rate density. The RMS error associated to the (1 − D)-field is equal to 0.08.

In the third step, the thermodynamic consistency is added. In the identification stage, only the data points for which $\dot{D} \geq 0$ and $\dot{Y} \geq 0$ are considered. It follows that for the first load step, all data are considered, for the second all but one, for the third all but nine, in the fourth 18 data are not considered and in the last 17. Figure 6 shows the new results. The power $\alpha$ is still equal to −0.37 and the RMS error is now reduced to 0.074. The same correction was performed on the (1 − D)-fields as in the previous analysis.

Lastly, the fourth step accounts for points for which the identification is deemed accurate enough according to the local damage identification error $\theta_e$. This analysis follows the previous one so that all data points are also thermodynamically consistent. A maximum value of 0.0225, i.e., about two times the average over all load levels. For the first three load levels, about 87 can be considered; for the last two only about 71 points are still admissible. In Figure 7, the new data are shown with the best fit. An RMS error of 0.07 is now achieved. It is believed that a part of the remaining scatter is related to the heterogeneous microstructure on the scale of the measurements. A power $\alpha$ is now equal to −0.39.

From the previous analysis, the (1 − D) maps can be plotted again by using the same multiplicative correction as in Figure 7. They are now consistent with the proposed growth law (33). Figure 8 shows the corrected $(1-D)$-field computed from the measured displacement field. Lower stiffnesses can be observed for the last load levels, thereby indicating higher damage states as for the first identification (Figure 2). The drawback of that type of approach is that the multiplicative constant (but just one) is still unknown. However, by assuming that for the first load level, the minimum value of damage is 0, gives an estimate of the latter. In the present case, the multiplicative constant in Equation (33) is equal to $A = 7.7 \times 10^{-3}$ with $\alpha = -0.39$.

## SUMMARY

An identification procedure is used to evaluate damage fields by using kinematic fields and then the damage growth law. The equilibrium gap method used herein is a non-standard finite element formulation in which the nodal displacements are known (i.e., measured in practice) and the elastic properties (or the damage field) are unknown. The latter are assumed to remain uniform over each element, but vary from element to element. When considering quadratic elements and only dealing with middle nodes, a linear system was derived in which the unknowns are written in logarithmic form. Such a procedure gives access to a space-varying field of elastic properties and/or damage fields.

The example of a cross-shaped specimen loaded along two perpendicular directions allowed us to analyze the damage state and changes prior to any visible discontinuity on the measured displacement field. Up to this point, the damage field is nothing but a simple way to account for a heterogeneous stiffness throughout the sample. The additional step that is proposed here, and tested against experimental data, is to require for an additional consistency, namely that the damage inhomogeneity results from a homogeneous damage law, combined with a heterogeneous loading. This is a strong statement, which was not used in the first damage field estimates. In a post-processing stage, it was shown that this additional consistency could be obeyed by a slight adjustment of the damage field based on a global scaling that remains undetermined in the present procedure. The fact that the damage field can be related to a unique function of its associated force is a very severe check of both the identification method and the homogeneous damage law hypothesis.

Progress in this problem can be envisioned along different directions, namely either by including the constitutive law hypothesis earlier in the identification procedure, so that the local stiffness determination makes use of the consistency assumption (e.g., Clausius-Duhem inequality), or by refining the post-processing stage so that, for instance, the quality of the algebraic form of the constitutive equation could be quantitatively measured, and hence could be used as a guide to complexify the damage law (e.g., anisotropic damage description, irreversible strain) to follow more closely the experimental results.

## Acknowledgments

This work was supported by CNRS in a project entitled "Analyses multi-échelles de champs de déformation par traitement d'image : vers l'identification de champs de propriétés mécaniques." The biaxial experiment reported herein was performed with the help of Dr. S. Calloch.

# REFERENCES

Allix, O. and F. Hild (Eds.). 2002. *Continuum Damage Mechanics of Materials and Structures*. Elsevier, Amsterdam (the Netherlands).

Andrieux, S., A. B. Abda and H. D. Bui (1997). *C. R. Acad. Sci. Paris*, Série I: 1431-1438.

Andrieux, S., A. B. Abda and H. D. Bui (1999). *Inverse Problems*, 15: 59-65.

Babushka, I. and W. C. Rheinboldt (1978a). *SIAM J. Num. Anal.*, 15: 736-754.

Babushka, I. and W. C. Rheinboldt (1978b). *Int. J. Num. Meth. Engng.*, 12: 1597-1615.

Bergonnier, S., F. Hild and S. Roux (2005). *J. Strain Analysis*, 40: 185-197.

Berthaud, Y., J.-M. Torrenti and C. Fond (1997). *Exp. Mech.*, 37: 216-220.

Bui, H. D. (1995). *Sur quelques problèmes inverses élastiques en mécanique de l'endommagement*. In Proc. *2e Colloque national de calcul des structures*, Hermes, Paris (France), 25-35.

Bui, H. D. and A. Constantinescu (2000). *Arch. Mech.*, 52: 511-522.

Burr, A., F. Hild and F. A. Leckie (1995). *Arch. Appl. Mech.*, 65: 437-456.

Calloch, S., D. Dureisseix and F. Hild (2002). *Technologies et Formations*, 100: 36-41.

Calloch, S., F. Hild, C. Doudard, C. Bouvet and C. Lexcellent (2001). *Analyse d'essais de compression biaxiale sur un A.M.F. à l'aide d'une technique d'intercorrélation d'images numériques*. In Proc. *Photomécanique 2001*, GAMAC, 207-214.

Chaboche, J.-L. (1977). *Sur l'utilisation des variables d'état interne pour la description du comportement viscoplatique et de la rupture par endommagement*. In Proc. *Problemes non-linéaires de mécanique*, Symposium franco-polonais, Cracow (Poland), 137-159.

Chevalier, L., S. Calloch, F. Hild and Y. Marco (2001). *Eur. J. Mech. A/Solids*, 20: 169-187.

Choi, D., J. L. Thorpe and R. Hanna (1991). *Wood Sci. Technol.*, 25: 251-262.

Claire, D., F. Hild and S. Roux (2002). *C. R. Mecanique*, 330: 729-734.

Claire, D., F. Hild and S. Roux (2004). *Int. J. Num. Meth. Engng.*, 61: 189-208.

Collin, F., Y. Berthaud and F. Hild (1998). *Visualisation par analyse d'images de la répartition des déformations et de l'amorçage dans un matériau composite*. In Proc. *Photomécanique 98*, GAMAC, 241-248.

Dawicke, D. S. and M. S. Sutton (1994). *Exp. Mech.*, 34: 357-368.

Desrues, J., J. Lanier and P. Stutz (1985). *Eng. Fract. Mech.*, 21: 251-262.

Fayolle, X. 2004. *Corimage : programme de pilotage d'essais asservis sur une jauge de deformation optique*. CNAM dissertation, CNAM Paris.

Forquin, P., L. Rota, Y. Charles and F. Hild (2004). *Int. J. Fract.*, 125: 171-187.

Germain, P., Q. S. Nguyen and P. Suquet (1983). *ASME J. Appl. Mech.*, 50: 1010-1020.

Geymonat, G., F. Hild and S. Pagano (2002). *C. R. Mecanique*, 330: 403-408.

Grédiac, M. (1989). *C. R. Acad Sci. Paris*, 309: 1-5.

Grédiac, M. (2004). *Composites: Part A*, 35: 751-761.

Grédiac, M., E. Toussaint and F. Pierron (2002). *C. R. Mecanique*, 330: 107-112.

G'Sell, C., J.-M. Hiver, A. Dahnoun and A. Souahi (1992). *J. Mat. Sci.*, 27: 5031-5039.

Halphen, B. and Q. S. Nguyen (1975). *J. Méc.*, 14: 39-63.

Hild, F., B. Raka, M. Baudequin, S. Roux and F. Cantelaube (2002). *Appl. Optics*, IP 41: 6815-6828.

Kohn, R. V. and B. D. Lowe (1988). *Math. Mod. Num. Ana.*, 22: 119-158.

Ladevèze, P. 1975. *Comparaison de modèles de milieux continus*. thèse d'Etat, Université Paris 6.

Lemaitre, J. 1992. *A Course on Damage Mechanics*. Springer-Verlag, Berlin (Germany).

Lemaitre, J. and J. Dufailly (1977). *Modélisation et identification de l'endommagement plastique des métaux*. In Proc. *3e congrès français de mécanique*, Grenoble (France).

Lemaitre, J. and J. Dufailly (1987). *Eng. Fract. Mech.*, 28: 643-661.

Lemaitre, J. and D. Marquis (1992). *ASME J. Eng. Mater. Tech.*, 114: 250-254.

Marigo, J.-J. (1981). *C. R. Acad. Sci. Paris*, t. 292: 1309-1312.

Press, W. H., S. A. Teukolsky, W. T. Vetterling and B. P. Flannery. 1992. *Numerical Recipes in Fortran*. Cambridge University Press, Cambridge (USA).

Rastogi, P. K. (ed.) 2000. *Photomechanics*. Springer, Berlin (Germany).

Roux, S., F. Hild and Y. Berthaud (2002). *Appl. Optics*, 41: 108-115.

Zienkiewicz, O. C. and R. L. Taylor. 1988. In *The Finite Element Method*. McGraw-Hill,, London (UK), 398-435.

# FIGURE CAPTIONS

Figure 1  a-Microstructure of the studied composite. b-View of the sample and region of interest (white box). c-Displacement field measured by digital image correlation for one load level (11 kN) close to failure (11.1 kN) in the region of interest.

Figure 2  Identified $(1 - D)$ fields for five load levels. In each case, the maximum value of $(1 - D)$ is set to 1.

Figure 3  Error indicator $\Theta$ versus load level. From the analysis of the results, it is expected that crack initiation occurred between 9 and 10 kN.

Figure 4  Change of $1 - D$ with $2Y / E_0$ for five different load levels. The symbols are identification points and the solid line is the best fit according to Equation (33). The values of the identified parameters are reported for each load level. For the sake of comparison, the same range is used for all load levels.

Figure 5  Change of $1 - D$ after correction with $2Y / E_0$ when all load levels are considered in one identification so that a single value for the power $\alpha$ is obtained. The symbols are identification points and the solid line is the best fit according to Equation (33).

Figure 6  Change of $1 - D$ after correction with $2Y / E_0$ when all load levels are considered and only the thermodynamically admissible data are kept. The symbols are identification points and the solid line is the best fit according to Equation (33).

Figure 7  Change of $1 - D$ after correction with $2Y / E_0$ when all load levels are considered. The thermodynamic consistency and identification accuracy are enforced. The symbols are identification points and the solid line is the best fit according to Equation (33).

Figure 8 Corrected $(1 - D)$-fields for five load levels. The maximum value of $1 - D$ is set to 1 for the lowest load level.

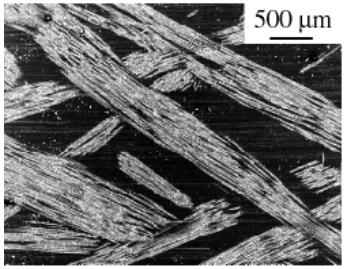 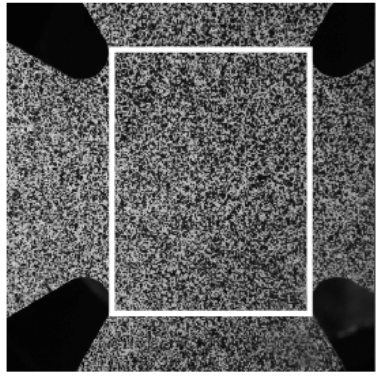 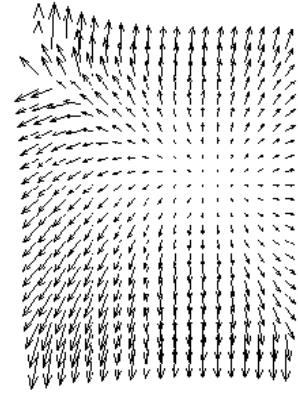

-a-     -b-     -c-

Figure 1. Claire *et al.*

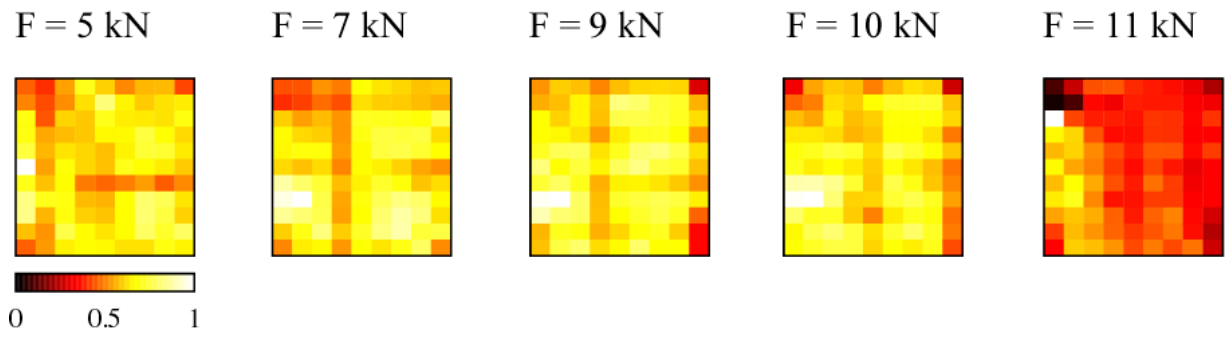

Figure 2. Claire *et al*.

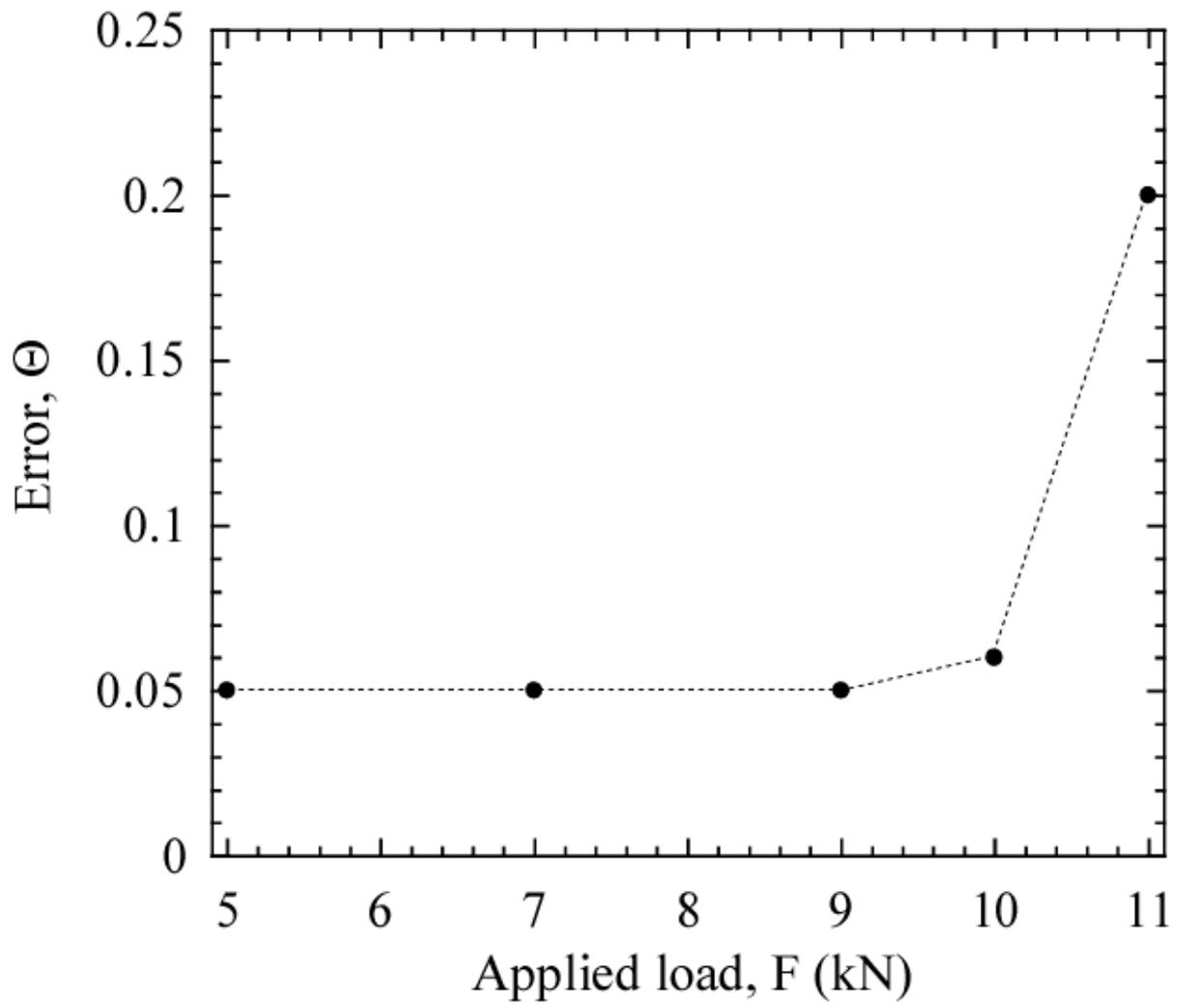

Figure 3. Claire *et al.*

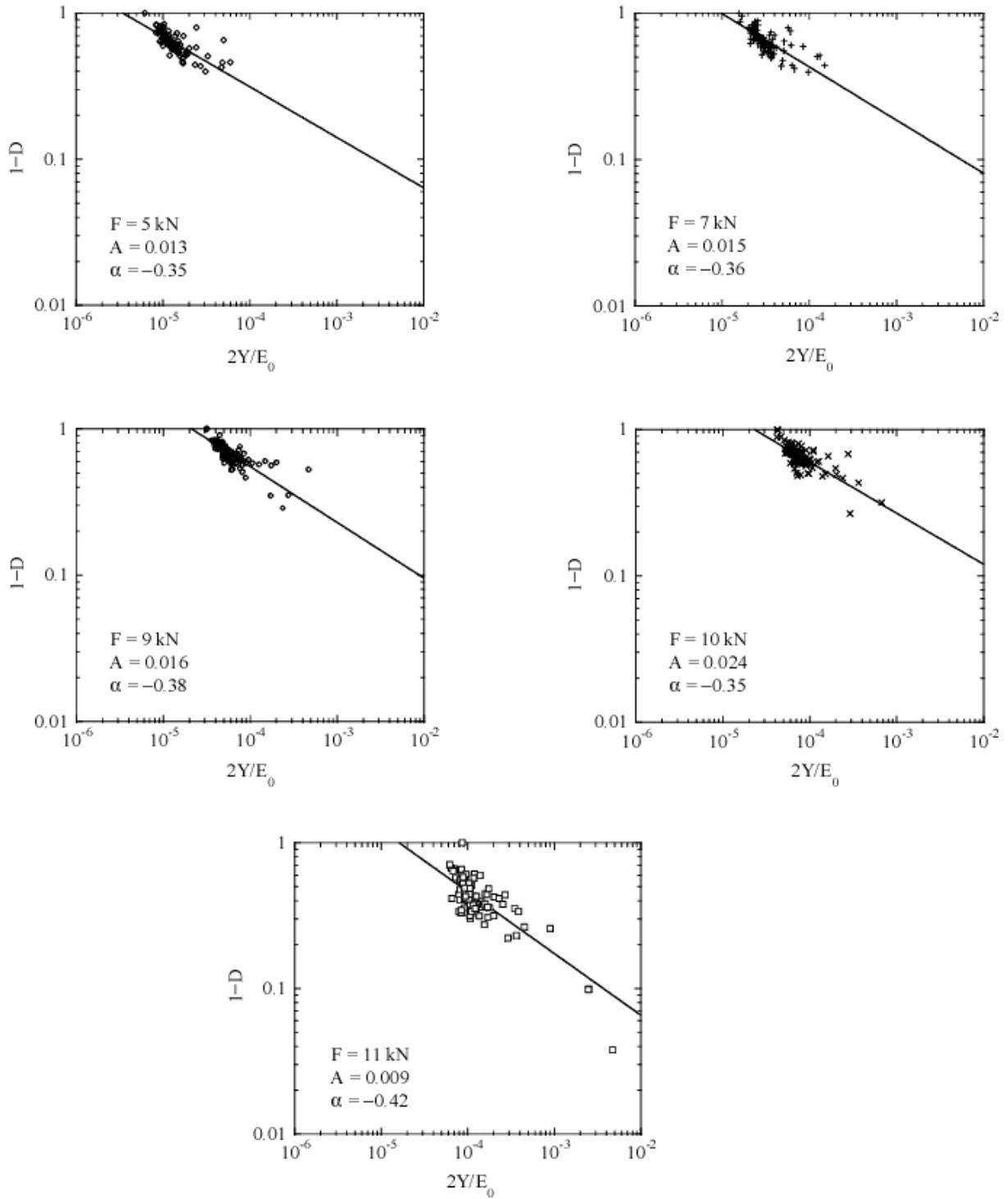

Figure 4. Claire *et al*.

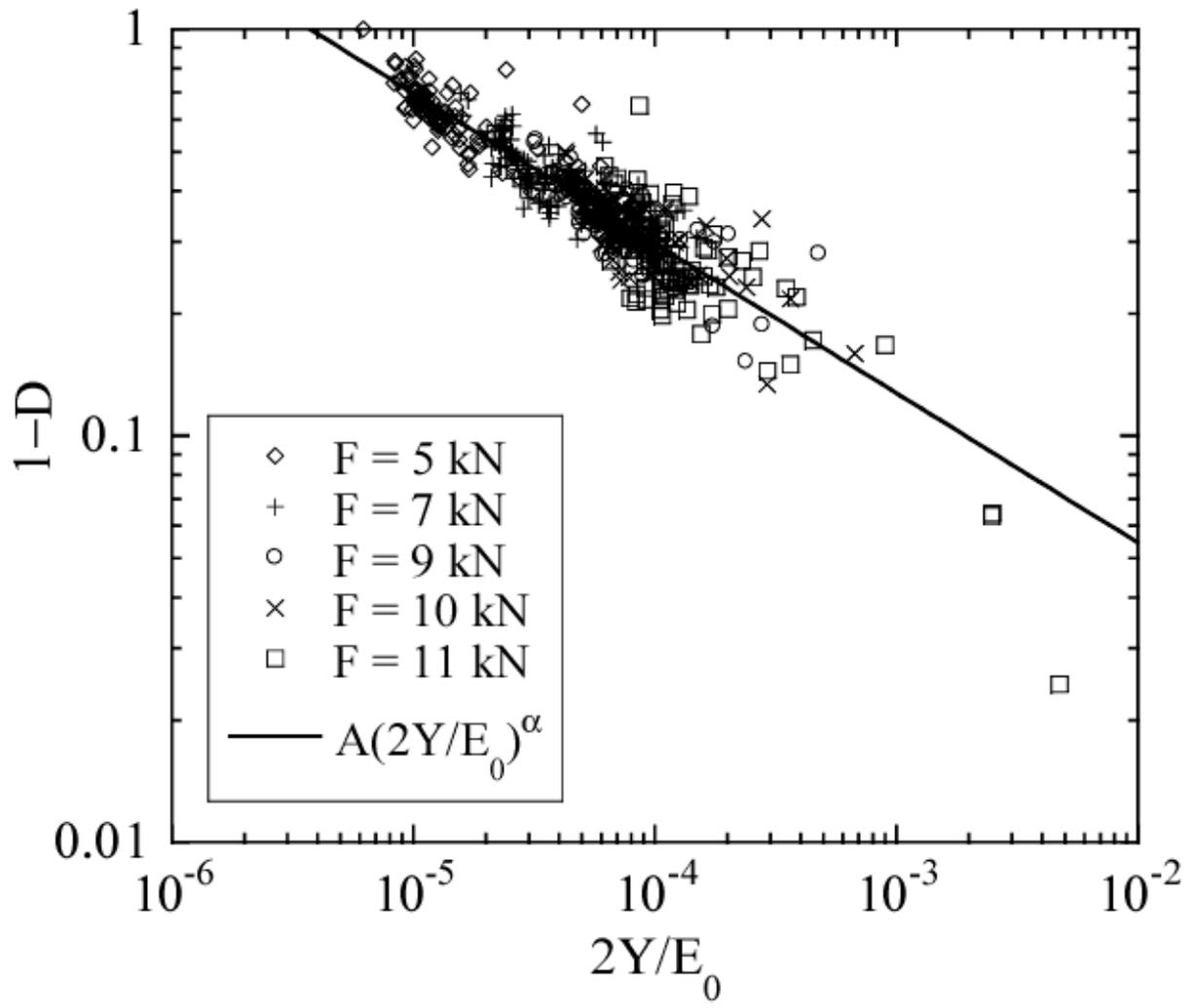

Figure 5. Claire *et al.*

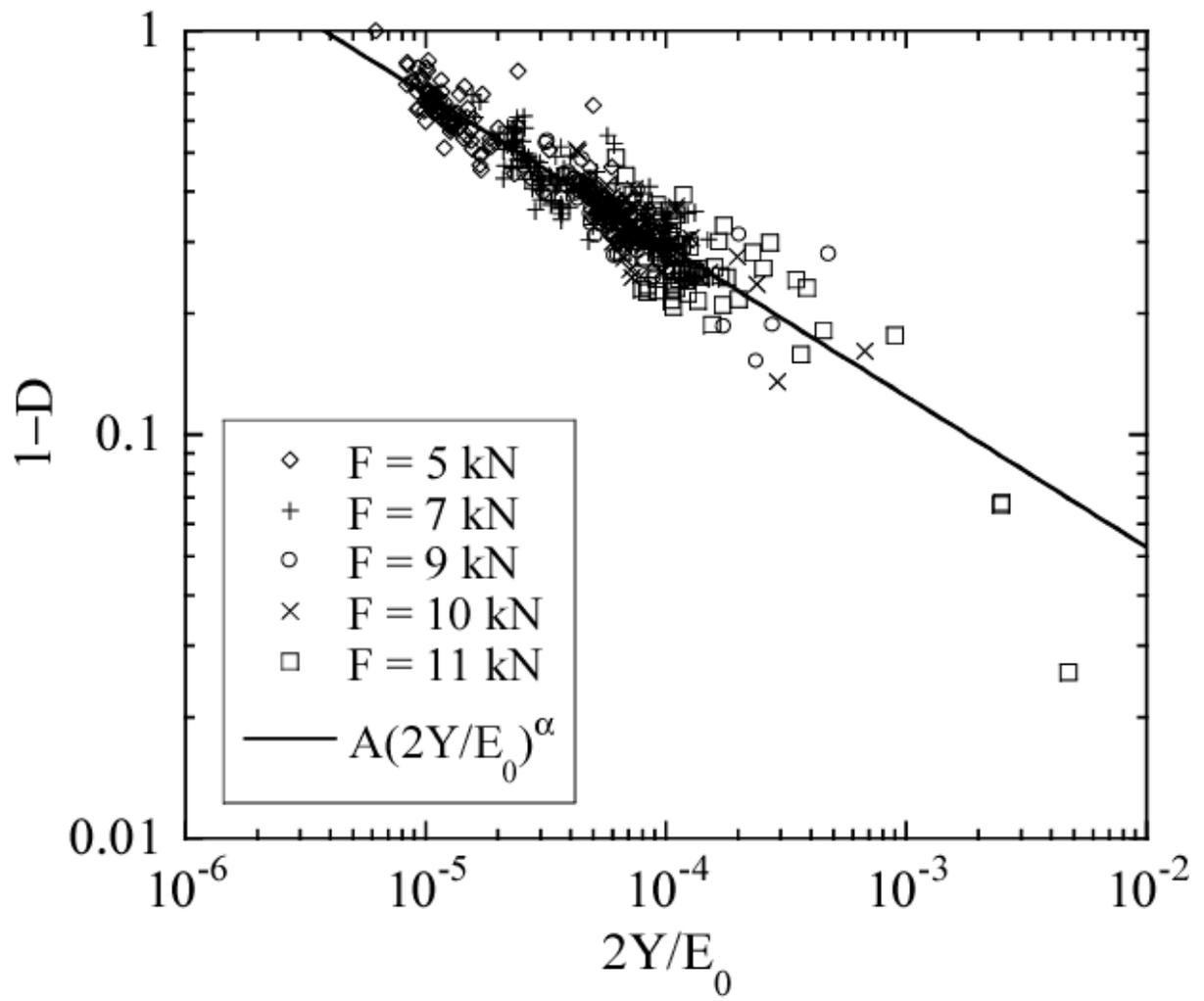

Figure 6. Claire *et al.*

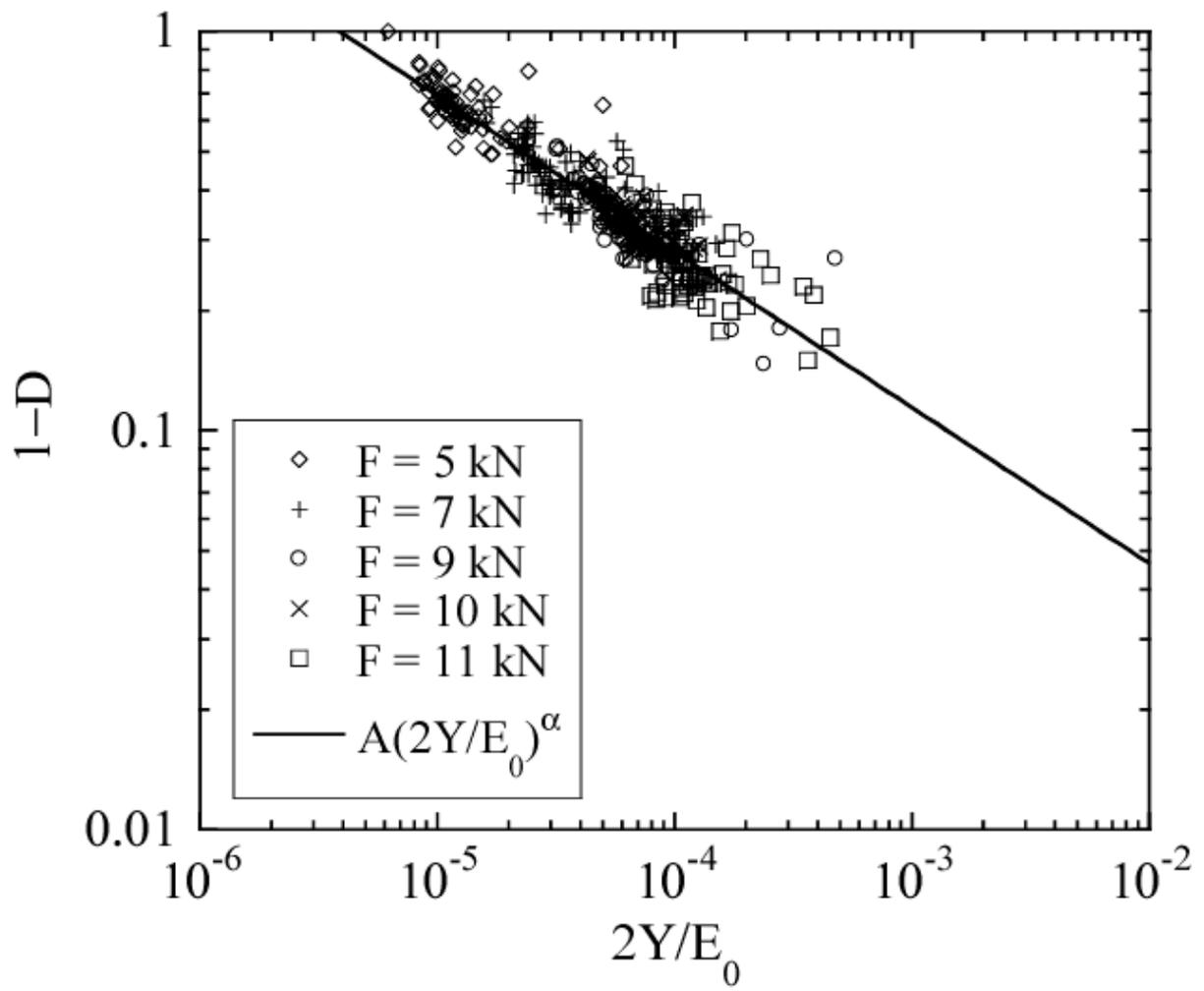

Figure 7. Claire *et al.*

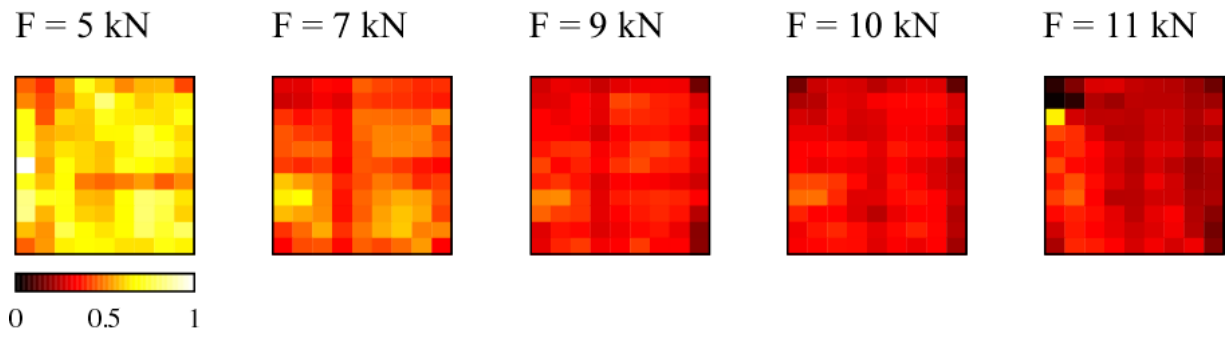

Figure 8. Claire *et al.*